# How big is the Sun: Solar diameter changes over time

J.P. Rozelot [1], A.G. Kosovichev [2], A. Kilcik [3]

[1] Université de la Côte d'Azur, Grasse, France
[2] New Jersey Institute of Technology, USA
[3] Akdeniz University, Antalya, Turkey

E mail (jp.rozelot@orange.fr).

**Abstract:** The measurement of the Sun's diameter has been first tackled by the Greek astronomers from a geometric point of view. Their estimation of ≈ 1800″, although incorrect, was not truly called into question for several centuries. The first pioneer works for measuring the Sun's diameter with an astrometric precision were made around the year 1660 by Gabriel Mouton, then by Picard and La Hire. A canonical value of the solar radius of 959".63 was adopted by Auwers in 1891. Despite considerable efforts during the second half of the XXth century, involving dedicated space instruments, no consensus was reached on this issue. However, with the advent of high sensitivity instruments on board satellites, such as the Michelson Doppler Imager (MDI) on Solar and Heliospheric Observatory (SoHO) and the Helioseismic and Magnetic Imager (HMI) aboard NASA's Solar Dynamics Observatory (SDO), it was possible to extract with an unprecedented accuracy the surface gravity oscillation $f$ modes, over nearly two solar cycles, from 1996 to 2017. Their analysis in the range of angular degree l = 140 - 300 shows that the so-called "seismic radius" exhibits a temporal variability in anti-phase with the solar activity. Even if the link between the two radii (photospheric and seismic) can be made only through modeling, such measurements provide an interesting alternative which led to a revision of the standard solar radius by the International Astronomical Union in 2015. This new look on such modern measurements of the Sun's global changes from 1996 to 2017 gives a new way for peering into the solar interior, mainly to better understand the subsurface fields which play an important role in the implementation of the solar cycles.

**Introduction.**

The first determinations of the diameter of the Sun have been made by the Greek astronomers through brilliant geometric procedures. Aristarchus of Samos (circa 310-230 BC), was able to set up the solar diameter $D_\odot$ as the 720th part of the zodiacal circle, or 1800 seconds of arc (″) (i.e. 360°/720). A few years later, Archimedes (circa 287-212 BC) wrote in the *Sand-reckoner* that the apparent diameter of the Sun appeared to lie between the 164th and the 200th part of the right angle, and so, the solar diameter $D_\odot$ could be estimated between 1620″ and 1976″ (or 27′00″ and 32′56″ [Lejeune, 1947; Shapiro, 1975]. Their results, albeit somewhat erroneous, were not truly called into question during several centuries. Such an issue is not only historical: in the quest to measure the solar diameter, important discoveries were made, such as its annual variation, which led to the eccentricity of the Earth's orbit by Ibn al Shatir from the Marāgha school (XIVth century). Another example is given by the analysis of some 53-year record observations of the solar diameter and sunspot positions during the seventeenth century, which lead to the conclusion that the solar diameter was larger and rotation slower during the Maunder minimum [Ribes et al. in 1987]. Moreover, through the irradiance variability, directly linked to the radius variability, its size may impact the temperature of the Earth [Eddy, 1979].

Indeed, the pioneer works for measuring the Sun's diameter were made around the year 1660 by Gabriel Mouton, immediately followed by Jean-Félix Picard and Philippe de La Hire. But, in spite of considerable efforts since the 17th century, which can be considered as the turning point of the astrometric estimate of the solar diameter, and the development of new instrumentation and techniques, both on the ground and in space, the scientific community is still working on an accepted worldwide definition and value of the solar diameter [ISSI forum, 2018]. Yet, a critical analysis of the measurements and errors deduced from observations of the Venus transits in 1874 and 1882, and through a network of heliometers which were observing the Sun during the years 1873-1886 in Germany at Breslau, Gotha-Strassburg, Göttingen, Berlin and Hamburg [Auwers, 1891] concluded that the diameter of the Sun could be set up at (1919.26 ± 0.10)″. Combined with the Astronomical Unit value commonly used at that time, i.e. 149 597 900 000 m, it turns out that the solar diameter was $R_\odot$ = 695 996.9 km. This 959″.63 value for the solar radius was referenced as a canonical one up to 2015. In 2002, the General Assembly of the International Astronomical Union in Pekin (CN) reconsidered the value of the Astronomical of Unit (AU), which was adopted as 149 597 870 700 m, and consequently the solar radius became 695 996.7 km, which is not significantly different. As a matter of historical curiosity, the Allen Astrophysical quantities, in the 1955 edition gives $R_\odot$ = (6.960 ± 0.001)×$10^{10}$ cm, the limb shape is circular to precision ± 0″.01, and the semi-diameter plus "irradiation" = 961″.2 (for observing the limb). The same Allen Astrophysical quantities, in the third edition (1973) states $R_\odot$ = 6.9599(7)×$10^{10}$ cm, still gives the semi diameter plus "irradiation", and substitutes the circular shape by oblateness with the equator-pole semi-difference = 0″.05. Lastly, the Allen Astrophysical quantities published in 2010 gives $R_\odot$ = (6.955 08 ± 0.00026)×$10^{10}$ cm, the oblateness of 0″.0086, and the "irradiation" correction disappears. This $R_\odot$ estimate was based on the determination made by [Brown and Christensen-Dalsgaard, 1998] who combined photoelectric measurements with models of the solar limb-darkening function. They further found that the annual averages of the radius are identical within the measurement error of ± 0.037 Mm.

However, in 2015 the International Astronomical Union General Assembly in Honolulu (USA) recommended that the solar radius $R^N$ *(nominal)* must be set at 6.957 $10^8$ m [IAU GA 2015]. As the AU was already determined, it turns out now that the angle subtended by the solar radius at the center of the Earth is 959″.22 and cannot be changed for the time being. This account for a difference of 0″.41 or 297 km in comparison with the Auwers value.

Figure 1 shows in the left panel the solar radius measurements made since the seventeenth century according to the survey of historical measurements compiled by [Rozelot and Damiani, 2012]. Space measurements made by the Michelson Doppler Imager (MDI) on Solar and Heliospheric Observatory (SoHO) [Scherrer et al., 1995], the Helioseismic and Magnetic Imager (HMI) aboard NASA's Solar Dynamics Observatory (SDO) [Scherrer et al., 2012], and the SODISM aboard CNES PICARD [Meftah et al., 2014], satellites have been added (in blue). The mean value of all these measurements is close to the canonical value of 959″.63. The right panel shows the solar radius measurements made since 1970 and leads to the same conclusion. Note that data can be found in [Wittmann, 1977], from 1836 to 1975, or [Djafer et al., 2008, Table 1] or [Kuhn et al., 2004]; the survey can be completed by the data listed in [Rozelot et al., 2016].

Figure 1 exhibits a great dispersion, spanning several tenths of arc-second, meaning that even with gradually increasing accuracy, by different means or techniques, including space measurements, the results obtained so far do not allow, even by means of a conscientious inspection, to determine without ambiguity a value of the solar diameter. All determinations made by different authors must be carefully analyzed by errors budget that have not always been made. Comparing each technique between them is not possible due, for instance, to the lack of calibration, stability evolution of the instrument with time, degradation of the filters, etc. However, one can see that the radius deduced from either space data, including the most recent analysis of eclipses data, seems systematically higher that the Auwer's canonical value, which raises interesting issues.

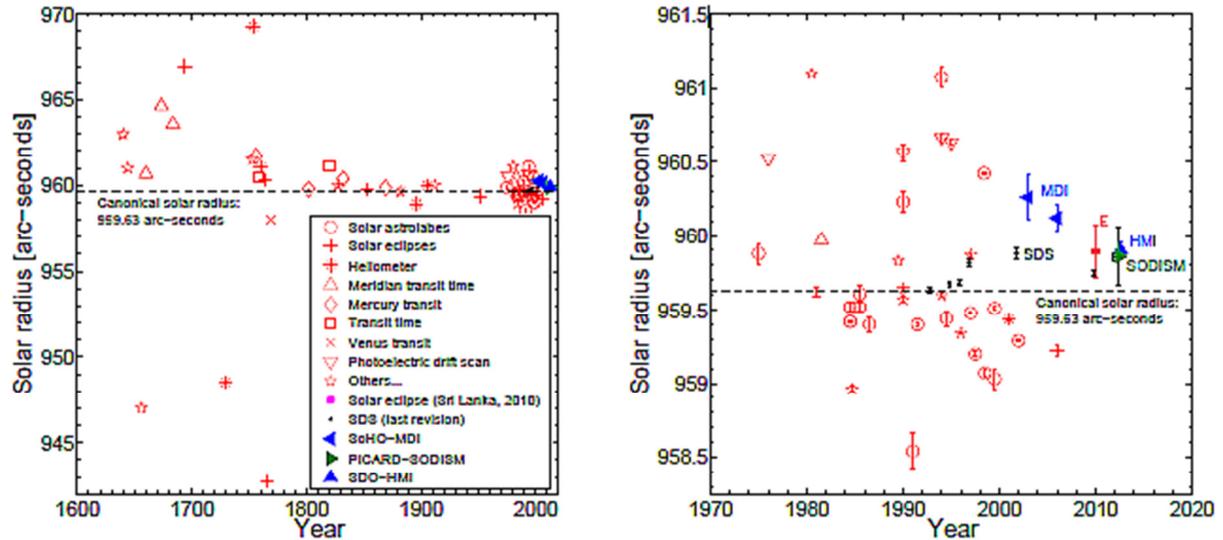

Figure 1. *(Left) Solar radius measurements (red symbols) made since the seventeenth century according to the survey of historical measurements made by [Rozelot and Damiani, 2012]. Blue symbols: space measurements. The mean value of all these measurements is about the canonical value of 950".63. (Right). Focus on the solar radius measurements made since 1970. Symbol E shows the solar eclipse determination in Sri Lanka (ground measurement) by [Adassuriya et al., 2011], i.e. (959.89±0.18)". Black circles: SDS balloon flight measurements [Sofia et al. 2013]. Blue symbols: space measurements from SoHO-MDI and SDO-HMI satellites. Green symbol: PICARD-SODISM measurement obtained during the transit of Venus (outside the atmosphere); according to [Meftah et al., 2014]. Note that the measurements were not carried out with the same type of instruments, in the same conditions (ground, space or solar eclipses) and were analyzed according different techniques, which complicates comparative assessments.*

Indeed, measuring the diameter of a gaseous sphere like the Sun is not an easy task. There are several diameter definitions; the most widely used is based on the determination of the inflection point of the limb profile, but others are permitted, such as the distance between two points on each limb that have an intensity equal to a given fraction of the intensity at the Sun's center, etc. These are the "photospheric" diameters determined from an optical point of view, at a given wavelength. The difficulty is that they do not lead, after analysis, to convergent values, and the way to link those needs physical models. A review of the different approaches is made for instance in Thuillier et al. [2005], Rozelot and Damiani [2012] or Meftah et al. [2014].

In the past, it was claimed that solar astrolabes permitted to emphasize a solar radius variation of about 0″.5 of amplitude with a 900 days periodicity. Such an intriguing result was the source of many commentaries and the outburst of series of other modified astrolabes built in particular in Rio de Janeiro (Brazil), São Paulo (Brazil), Santiago (Chile), Antalaya (Turkey) and Tamanrasset (Algeria). Unfortunately, in spite of efforts aiming at standardizing procedures and software, astrolabes results remain conflicting, both in amplitude and in phase, and this network was progressively abandoned. However [Andrei et al., 2015], analyzing 42 years of continuous observations of the solar diameter from 1974 to 2015 found a mean estimate of 959″.94, evidencing a change by comparing the Auwers' radius of 959″.63. The authors emphasize the importance of "keeping and making available such long, continuous, and uniform series of solar diameter measurements, to ensure a homogenous and continuous monitor of the solar diameter without the lifetime limitation of the satellites".

Ground-based observations suffer from the degradation of the solar signal by the atmospheric turbulence, which require corrections. Many works have been devoted to this subject, the main issue being to know how to deconvolute the observed data by an adequate function which expresses the state of the atmosphere at the time of the observation. One can see for instance [Hill et al., 1975, Rozelot et al., 2003 or Irbah et al., 2003]. Thanks to the new instrumentations at the Calern Observatory (South France) which will permit to observe in real time both the solar and the atmospheric signals, it will be possible in a near future to extend the series obtained from space. Indeed, the observations made outside the atmosphere are in principle, more favorable. Let us recall the first observations made by means of balloon flights (SDS -Solar Disk Sextant- experiment from S. Sofia et al. [Maier 1992; Sofia et al., 2013], and the most recent ones made through the PICARD satellite (Meftah et al., 2015), leading to a solar radius between 959″.76 and 959″.86. These are the only two results where a realistic errors budget is proposed for an accurate determination of the "photospheric" solar radius (+/-0.12 arcseconds for SDS and +/-0.19 arc-second for PICARD).

Determination of the solar diameter through eclipses has been done also since long ago as they present some decisive advantages, the main one being due to the occultation that takes place in space, and so is free from atmospheric effects [Lamy et al., 2015]. The first possibility to determine the solar radius was made on the basis of the annular eclipse observed by Clavius in 1567. This eclipse with the standard solar radius should have been total, while the solar photosphere exceeded the lunar limb to show the observed ring. Under this exceptional circumstance, the diameter of the Sun was exactly the diameter of the Moon [Stephenson et al., 1997], and thus can be computed. A survey of solar radius measurements made during eclipses has been made in [Damiani et al., 2012, Fig. 2.] and can be complemented with the work of [Lamy et al., 2015] who found a mean radius of (959.99 ± 0.06)″ or (696 258 ± 44) km during the eclipses of 2010 [(959.94 ± 0.02)″], 2012 [(960.02 ± 0.04)″], 2013 [(959.99 ± 0.09)″], and 2015 [(960.01 ± 0.09)″. In these experiments, the radius was determined through an innovative method based on light curves recorded just before and after second and third contacts, through a minimization process, and performed by pre-programmed photometers at 540 nm. The above-mentioned results contribute in suspecting that the former photospheric canonical value (959″.63) was too low. Correcting the 540-nm value of 959″.99 arcsec by about 0″.46 for the apparent larger radius when observing the disc, it is obtained 959″.53, which is still 0″.3 above the IAU 2015 adopted value.

**Helioseismology** can provide an interesting alternative. With accurate measurement of f-mode frequencies by the GONG (Global Oscillations Network Group) [Harvey et al., 1996] and by the SoHO/MDI [Scherrer et al., 1992] and SDO/HMI space instrument [Scherrer et al., 2012], it has been shown that global helioseismology provides important information about the physics of localized structures beneath the surface of the Sun, and it has been suggested also to use the f modes to estimate the solar radius, which is thus called the "seismic radius". The procedure has been described by [Schou et al. 1997] and explained in detail in [Dziembowski et al., 2001] and in [Di Mauro, 2003]. Results lead to a seismic radius less than the photospheric one, reduced by approximately 300 km, and it is this value which was adopted by the IAU GA in Honolulu (USA) in 2015, as seen above.

The depth at which f-modes are trapped is located at about 1 Mm, for $\ell = 300$, while the lower limit of $\ell = 140$ extends to a depth of about 12 Mm. Values of the radius obtained by such a way are consequently smaller than the photospheric one. However, the f-modes are not trapped between a pair of rigid boundaries and the extent of region covered by them will depend on the definition of boundary as well as on the precise mechanism responsible for frequency variations. Defining the photospheric radius to be where the Rosseland optical depth is $\tau = 2/3$, the difference between the two radii account to around 300 km (0″.4). This apparent discrepancy between the "photospheric" and the "seismic" radius was solved by [Habereitter et al., 2008] from radiative transfer calculations with the help of a code named COSI (Code for Solar Irradiance, a combination of a model atmosphere code in spherical symmetry and the spectrum synthesis program SYNSPEC, adapted to the solar atmosphere). A discussion of the different models to derive the solar limb profile can be found in Thuillier et al. [2011].

If there is no consensus today, this could be due to several reasons: (i) there is no unique method of estimation of the Sun's radius, as just said above (some observers define the radius as the distance to the point of inflexion in the limb profile, whilst the theoretical solar models assume the solar surface to be located at a point where the optical depth is of order unity), (ii) instrumental effects (point-spread function, spatial resolution, optical distortion, temperature variations) not properly taken into account, (iii) atmospheric turbulence on the observed Sun's full disk pictures or on the Sun's limb, which are not corrected, (iv) different instruments observing at different phases of the Solar Cycle and, (v) observations not made at the same wavelengths. As regards to this last issue, certain instruments observed in the continuum at different wavelengths, while others observed in the center of a Fraunhofer line, as was done for example at the Mount Wilson observatory (USA). In addition, there are instruments that used a narrow band pass, such as MDI on board SoHO (0.0094 nm) and Mount Wilson (0.014 nm), whereas others used a wide spectral domain on the order of hundreds of nanometers, as it was done by the CCD astrolabes. Moreover, the errors deduced from each technique are obtained from the dispersion of the measurements and do not include systematic effects, that have to be evaluated and added. It results that the differences in the absolute values of the radius can be larger than the quoted errors obtained by each experiment. In such a complex situation, comparison of the results as regards their eventually diverging views and reasons is cardinal. Probably, only space-based measurements could provide an accurate determination of the solar radius with a realistic errors budget.

**2. How large are the temporal variations of the solar diameter?**

The temporal dependence of the solar diameter is a long-standing question. The turning point was made in 1874 when it was argued that the changes were proportional to the inverse of the solar activity (the so-called Secchi-Rosa [1874] law). Observational determinations lead up-to-now to conflicted results.

- On *mid-term variation*s, i.e. along two or three solar cycles, at least four groups of observers have claimed that the solar radius varies in phase with surface activity, seven groups of observers have reported radius changes in antiphase with surface activity, while four groups of observers have reported no significant change at all [Stothers, 2006, and references herein]. From space [Kuhn et al., 2004] have reported that during Solar Cycle 23, between the solar minimum and solar maximum, the radius of the Sun did not change by more than ± 7 mas (14 mas peak to peak, i.e. ≈ 10 km).

- On *long term variations* (secular trends) [Eddy, 1978], analyzing daily meridian transit timings of the Sun made at the Royal Greenwich Observatory (UK) from 1836 to 1953 evidenced a secular decrease of the solar diameter of about 2.25 arsec/century, A few years later, an analysis of observations of 23 transits of Mercury in front of the Sun between 1736 and 1973 by [Shapiro, 1980] have shown no indication of any significant change in the diameter of the Sun yielding a decrease of the angular diameter, as viewed from the Earth, of under 0.3 arc second per century, incompatible with the value obtained by Eddy. [Gilliland, 1981] analyzing five different data sets, concluded that a secular decrease of ≈ 0″.1 per century over the last 265 years is likely. Other works from [Wittmann, 1980, Stephenson et al., 1997, Parkinson, 1988], did not support the assumption of a secular decrease. Analyzing the diameter of the Sun over the past three centuries [Toulmonde, 1997] concluded that even by taking into account the instrumental corrections and refraction, the homogenized database does not reveal any sensible secular variation, limited to a 0″.1 uncertainty centered on 960″.0. By contrast, analyzing the same data set [Rozelot, 2006] evidenced again a greater Sun during the Maunder Minimum, but also a general declining trend of about 0″.5 over these three centuries. [Gallego et al., 2015] reported that the solar semi-diameter observed at the Cadiz (SP) Naval observatory from 1773 to 2006, do not present any significant trend in the past 250 years, or if any should be inside the 1″.77 measurements uncertainty. An analysis of the Kodaikanal (IN) daily digitized white light solar pictures, from 1923 to 2011, shows an apparent decrease of the solar radius of less than (5.8 ± 3.7) mas per year [Hiremath et al., 2018].

- On *theoretical grounds*, [Callebaut et al., 2002] were certainly the first to point out that changes in solar gravitational energy, in the solar upper layers, necessarily involve variations in the size of the envelope. The mechanism is simple. Bearing in mind the definition of the energy Eg = –∫(G$m/r$) d$m$, (where $r$ is the radial coordinate and G the gravitation constant) and assuming hydrostatic equilibrium, a thin shell of radius d$r$ containing a mass d$m$ in equilibrium under gravitational and pressure gradient forces will be expanded or contracted if any perturbation of these forces occurs. In [Fazel et al., 2008], the authors improved the method and show that any variations of the size of the solar envelope must be within a few km of amplitude over a Solar Cycle, a value in perfect agreement with those deduced from inversion of the f-modes in helioseismology.

The study *of the solar oscillation f modes* provided by the two space missions SoHO (Solar and Heliospheric Observatory) and SDO [Solar Dynamics Observatory] from 1996 to 2017 has been made by Kosovichev and Rozelot. [2017] in the scope to analyze the coefficients of rotational frequency splitting -that measure the latitudinal differential rotation-, together with the asphericity coefficients -that gives the profile of the limb shape over the heliographic latitudes- [Reiter et al., 2015]. As above mentioned, the f mode frequencies permit also to extract the seismic radius and its variations with the solar cycle. The analysis is focused on the low-frequency medium-degree f-modes observed in the range of $\ell$ = 140-300, where the kinetic energy is concentrated within a layer of approximately 15 Mm deep. The properties of these modes are affected by the surface magnetism and temperature/sound-speed changes, but also reflect large-scale variations in the near-surface shear layer [NSSL]. This layer (called

"leptocline" [Godier and Rozelot, 2001]) presumably plays an important role in the solar dynamo [Pipin and Kosovichev, 2011].

The f-mode frequency variations, $\langle\Delta\nu/\nu\rangle$, averaged for the whole common subset of modes in the whole observed angular degree range, $\ell$ = 137-299 have been calculated for 72-day periods. The quantity $-2/3\langle\Delta\nu/\nu\rangle$ representing a proxy of the relative seismic radius variations, $\langle\Delta R_{seis}/R_{seis}\rangle$ is displayed in Figure 2a. It exhibits a modulation with amplitude of about $2.3\times10^{-5}$ in Solar Cycle 23 and about $1.2\times10^{-5}$ in Cycle 24. Taking into account the Auwer's canonical value for the solar radius, such variations account respectively for 22 and 12 mas. Such a result was already pointed out by Fazel et al. [2008] on theoretical grounds, and also by Meftah et al. [2015]. It is clear that the seismic solar radius varies in phase opposition with the solar cycle (Figure 2b) with a very small-time lag which has been found at (0.044 ± 0.019) yr (correlation coefficient 0.94). When the averaging includes the f-modes of $\ell$ = 140-200, the seismic radius modulations are as twice as small, but the phase relation remains. As earlier pointed out by Lefebvre and Kosovichev [2005] and Lefebvre et al. [2007], the seismic radius variations are not constant and non-homologous in this subsurface layer: their examination requires a careful helioseismic inversion procedure, and will give us a real insight into changes of the solar stratification.

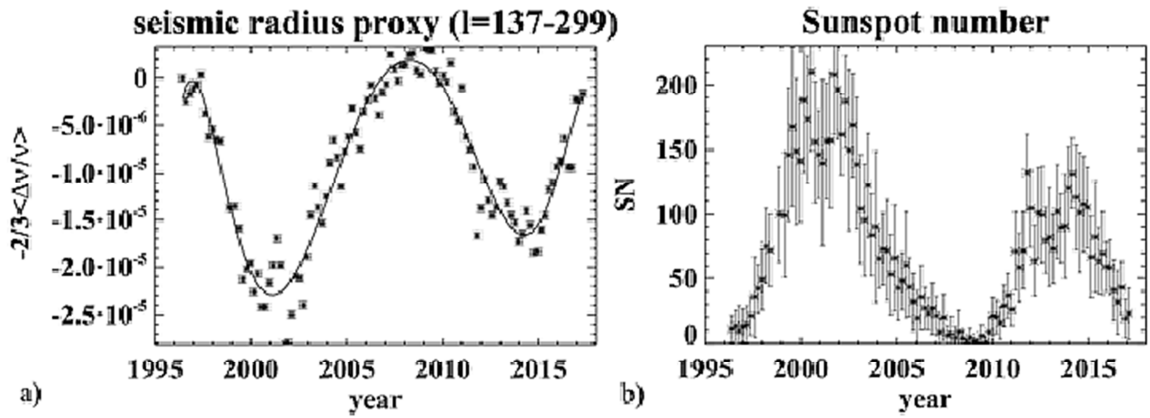

Figure 2. *a) Variation of the seismic radius proxy, $\langle\Delta R_{seis}/R_{seis}\rangle=-2/3\langle\Delta\nu/\nu\rangle$, relative to the first measurement in 1996, as deduced from the analysis of the f mode frequencies extracted from the MDI and HMI data from 1996 to 2017. The solid curve shows a polynomial fit. b) The sunspot number, SN (daily numbers in grey), averaged for the 72-day periods corresponding to the intervals of the helioseismic analysis (black dots). (After [Kosovichev and Rozelot, 2017]).*

**Conclusion**.
Due to the progress achieved these last years in high angular resolution and in the precise determination of ephemerides, it would be believed that the size of the Sun (and its deviation from a perfect sphere) could be known at a very high precision, or, at least, as well known as stellar radii obtained nowadays by interferometry. In spite of sophisticated techniques, the direct measure of the size of the Sun and its possible temporal variations, both on short term and secular variations remain vitiated by significant errors of measurements, both at the ground level (mainly due to atmospheric turbulence) and in space (limb shape displacement). A statistical analysis allows in principle to differentiate between different scenarios, but such a study has not been achieved so far in the absence of relevant data over a sufficient time span. To clarify the discrepancies which appear in this study, a much more robust comparison between solar radius measurements (mainly from space) and the IAU 2015 adopted value (that may include a normalized wavelength) remains to be done.

Helioseismology data obtained for nearly two solar cycles of observations from space provide a powerful tool for determining the variability of the seismic solar radius. The results show that the solar subsurface layer is shrinking or expanding not at the same level during the successive solar cycles. This may help to peer inside the solar interior and permit to investigate other important properties of the subsurface dynamics of the Sun, which previously were not accessible. However, the helioseismology data on their own do not provide variations of the photospheric radius. Further understanding of the solar radius variations will require detailed theoretical modeling of the structure and dynamics of the leptocline.

**Acknowledgment.**
The authors thank Katya Georgieva for her kind invitation to the Sunny Beach Summer School in Bulgaria. One of us (JPR) thanks also the International Space Science Institute (ISSI) in Bern (CH) where he is repeatedly invited as a visitor scientist.